\documentclass[aps,prb,nofootinbib,twocolumn,amsmath,amssymb,superscriptaddress,notitlepage]{revtex4-2}
\usepackage{latexsym}
\usepackage{graphicx}
\usepackage{times,psfrag,subfigure}
\usepackage{enumerate}
\usepackage{amsmath}
\usepackage{amsthm}
\usepackage{dsfont}
\usepackage{dcolumn}
\usepackage{bm,bbm}
\usepackage{wasysym}
\usepackage[normalem]{ulem} 
\usepackage{color}
\usepackage[dvipsnames]{xcolor}
\usepackage[breaklinks,colorlinks,allcolors=blue]{hyperref}
\usepackage{latexsym,amsmath,amssymb,bm,euscript}
\usepackage{textcomp}
\usepackage{tabularx}
\usepackage{multirow}
\usepackage{setspace}
\usepackage{ctable}
\usepackage{sidecap}
\usepackage{placeins}
\usepackage{threeparttable}
\usepackage{braket}
\usepackage{multirow}
\usepackage{comment}
\usepackage{float}
\usepackage{framed}
\usepackage{scalerel,stackengine}
\stackMath
\usepackage{booktabs}

\renewcommand{\arraystretch}{1.2} % row height
\definecolor{Qicolor}{RGB}{3, 136, 252}

\usepackage{ulem}

\let \oldbm \bm
\renewcommand{\vec}[1]{\oldbm{#1}}

\def\H{\mathcal{H}}
\def\T{\mathcal{T}}

\def\L{\mathcal{L}}

\def\P{\mathcal{P}}
\def\bbI{\mathbb{I}}

\newcommand{\sket}[1]{|#1\rangle\!\rangle}
\newcommand{\sbra}[1]{\langle\!\langle #1 |}

\newcommand\reallywidehat[1]{%
\savestack{\tmpbox}{\stretchto{%
  \scaleto{%
    \scalerel*[\widthof{\ensuremath{#1}}]{\kern-.6pt\bigwedge\kern-.6pt}%
    {\rule[-\textheight/2]{1ex}{\textheight}}%WIDTH-LIMITED BIG WEDGE
  }{\textheight}% 
}{0.5ex}}%
\stackon[1pt]{#1}{\tmpbox}%
}

\makeatletter
\renewcommand*\env@matrix[1][*\c@MaxMatrixCols c]{%
  \hskip -\arraycolsep
  \let\@ifnextchar\new@ifnextchar
  \array{#1}}
  
\makeatother

\allowdisplaybreaks

\graphicspath{{figures/}}

\DeclareMathOperator{\Tr}{Tr}

\makeatletter
\newcommand*{\balancecolsandclearpage}{%
  \close@column@grid
  \clearpage
}
\makeatother

\begin{document}
\title{Dissipative Yao-Lee Spin-Orbital Model: Exact Solvability and $\P\T$ Symmetry Breaking}

\author{Zihao Qi}
\email{zq73@cornell.edu}
\affiliation{Department of Physics, Cornell University, Ithaca, New York 14853, USA.}

\author{Yuan Xue}
\email{yuan\_xue@utexas.edu}
\affiliation{Department of Physics, The University of Texas at Austin, Austin, Texas 78712, USA.}

\date{\today}

\begin{abstract}
Exactly solvable dissipative models provide an analytical tool for studying the relaxation dynamics in open quantum systems. In this work, we study an exactly solvable model based on an anisotropic variant of the Yao-Lee spin-orbital model, with dissipation acting in the spin sector. We map Liouvillian dynamics to fermions hopping in a doubled Hilbert space under a non-Hermitian Hamiltonian and demonstrate the model's exact solvability. We analyze the model's strong and weak symmetries, which protect an exponentially large manifold of non-equilibrium steady states, establishing the system as a physically feasible dissipative spin liquid. Furthermore, we analyze the transient dynamics in a translationally invariant sector and discover that the single-particle Liouvillian spectrum hosts an exceptional ring in momentum space. We map out a characteristic $\P\T$ symmetry breaking transition driven by the dissipation strength, which governs the crossover from oscillatory to decaying relaxation of physical observables. Our work provides a physically motivated, solvable setting for exploring the coexistence of dissipative spin liquid physics and Liouvillian spectral singularities.
\end{abstract}

\maketitle

\tableofcontents

\section{Introduction}
The non-equilibrium dynamics of quantum many-body systems coupled to an environment, through decoherence, dissipation, or measurements, has emerged as a rapidly growing frontier in condensed matter physics~\cite{open1, open2, open3, measurement3, weakmeasurement1, open4, open5, measurement2}. While the coupling to an external environment has traditionally been viewed as a source of detrimental noise, recent developments have established dissipation as a powerful resource, particularly with advances in experimental control across atomic, optical, solid-state, and superconducting platforms~\cite{np_Diehl2008, np_DallaTorre2010,np_Barreiro2011,np_Eisert2015,np_Skou2021,Özdemir2019, IBMQ}. These developments have stimulated substantial interest in understanding how system-environment couplings can be engineered to shape relaxation dynamics, steady-state structure, and realize topological phenomena in driven-dissipative settings~\cite{cenke_spinliquid, decoherence_lee, driven1, driven2, driven3,driven4, dissipativeengineering5, dissipativeengineering6, dissipativeengineer2, dissipativeengineer3, dissipativeengineer4}.

Despite recent progress, the theoretical study of open many-body quantum systems remains challenging due the exponential explosion of the operator space. Compared to closed quantum systems, where the Hilbert space dimension scales as $2^L$ for a system with $L$ qubits, open quantum systems require treating density matrices whose dimension scales as $4^L$. This scaling quickly renders exact numerical simulations unfeasible beyond small system sizes. Tensor-network-based approaches~\cite{TN_PhysRevLett.93.207204,TN_PhysRevLett.93.207205,TN_PhysRevLett.114.220601,TN_PhysRevLett.116.237201,TN_PhysRevLett.132.200403,TN_Jaschke_2019,TN_hmbj-lg7p} and quantum-trajectory methods~\cite{traj_Daley04032014,traj_PhysRevA.45.4879,traj_PhysRevLett.68.580,traj_RevModPhys.70.101,traj_xue2025simulationbilayerhamiltoniansbased} have extended our reach, but they remain numerically costly and may obscure the underlying physics. In contrast, exactly solvable Liouvillians~\cite{BCS_solvable, Shackleton_dissipative, arovas_dissipative, lucas_sa, exact1, exact2,exact3,exact4, exact5, exact6, exact7} provide non-perturbative insight into the complex behavior of relaxation and steady states, and constitute valuable benchmarks for new numerical techniques.

Recently, a notable development along this line of inquiry is the dissipative spin liquids (DSL): exactly solvable Liouvillians with an extensive number of symmetries, which protect an exponentially large manifold of non-equilibrium steady states~\cite{Shackleton_dissipative, arovas_dissipative,lucas_sa}. 
However, existing DSL constructions are formulated in terms of Gamma matrices acting on enlarged on-site Hilbert spaces and lack a direct mapping to local spin degrees of freedom. This makes it difficult to assess how such models might be connected to realistic materials or solid-state platforms.

In this work, we introduce an exactly solvable dissipative model that is both physically motivated and dynamically rich. 
We consider an anisotropic variant of the Yao-Lee spin-orbital model~\cite{yaolee_original}, a well-studied generalization of Kitaev's honeycomb model~\cite{Kitaev_honeycomb} that incorporates orbital (pseudospin) degrees of freedom while retaining its exact solvability. The Yao-Lee model hosts a variety of interesting phenomena, such as spin-1 fermionic excitations, vison crystallization, and topological band transitions in the $\sigma$ sector~\cite{yaolee_original, yaolee1, yaolee2, yaolee3}. More recently, Ref.~\cite{yaolee_experiment} presents a road-map to realizing the Yao-Lee model in experiments. While non-Hermitian extension of the model has been studied~\cite{nonhermitian_yaolee}, how Lindbladian dissipation affects its spin-orbital dynamics and steady state structure has not been explored.

We solve the dissipative model through third quantization~\cite{third_quantization}, mapping the Liouvillian super-operator to a non-Hermitian Hamiltonian that is quadratic in fermions hopping on a bilayer honeycomb lattice. This mapping allows for an analysis of the system's strong and weak symmetries, from which we obtain a systematic criterion for determining flux sectors that contain non-equilibrium steady states (NESS). We identify an exponentially large manifold of NESS, which confirms that the model realizes a DSL~\cite{Shackleton_dissipative, arovas_dissipative, lucas_sa}, but in a setting where the local degrees of freedom are more closely connected to experiments.

While the NESS manifold characterizes the long-time limit, the relaxation to equilibrium contains rich transient dynamics. To explore this, we focus on a translationally invariant flux sector and compute the single-particle Liouvillian spectrum analytically. We find that the single-particle spectrum hosts an exceptional ring in momentum space: a continuous contour of exceptional points (EPs), where eigenvalues and eigenvectors coalesce~\cite{Huber_PT, prosen_PT,Nakanishi_PT, dissipative_stark, quantum_quench_drivendissipative, localizer}. 

As the dissipation strength increases, the system undergoes a characteristic sequence of $\P\T$ symmetry breaking transitions. $\P\T$ symmetry is preserved at weak dissipation: up to an overall shift, all single-particle Liouvillian modes have purely imaginary eigenvalues, leading to oscillatory dynamics. Crossing the exceptional ring produces a $\P\T$ mixed regime in which modes with imaginary and real eigenvalues coexist, before the spectrum becomes fully $\P\T$-broken and dynamics are governed by real decay rates. This behavior leads to a clear separation between two dynamical regimes, in which observables relax with different rates. Our work therefore provides a physically motivated, exactly solvable setting in which dissipative spin liquid physics and $\P\T$ symmetry breaking coexist in different symmetry sectors and timescales.

The rest of the paper is organized as follows. In Sec.~\ref{Sec: mapping}, we review the mapping from the Lindbladian super-operator to a non-Hermitian Hamiltonian in the doubled Hilbert space. In Sec.~\ref{Sec: exactly_solvable}, we introduce our dissipative spin-orbital model and demonstrate its exact solvability. Sec.~\ref{Sec: symmetry} details the model's extensive number of symmetries and the construction of the NESS manifold. In Sec.~\ref{sec:PT}, we analytically compute the Liouvillian spectrum in a specific symmetry sector, revealing a ring of exceptional points and the $\P\T$ symmetry breaking transition. We conclude in Sec.~\ref{sec:conclusion} with a summary and outlook.

\section{Mapping between Lindbladians and bilayer Hamiltonians \label{Sec: mapping}}
We first briefly review Lindbladian dynamics and the mapping from Lindbladians to non-Hermitian Hamiltonians on a doubled Hilbert space. Given a Hamiltonian $H$ and a set of Markovian jump operators $\{L_i\}$, the density matrix $\rho$ of an open quantum system evolves according to the Lindblad master equation~\cite{lindblad,Gorini}:
\begin{equation}
    \L[\rho] = -i[H,\rho] + \sum_i L_i\rho L_i^\dagger - \frac{1}{2}\{L_i^\dagger L_i, \rho\},
    \label{eq:lindbladian_eq}
\end{equation}
where $[\cdot, \cdot]$ and $\{ \cdot, \cdot\}$ denote the commutator and anti-commutator, respectively.

The Liouvillian $\mathcal{L}$ is the generator of a quantum dynamical semigroup, mapping between bounded operators on the original Hilbert space, $\mathcal{L}:B(\mathcal{H}) \to B(\mathcal{H})$.
If we treat the space of operators $B(\mathcal{H})$ as a Hilbert space by itself, equipped with the inner product $\langle \rho, \sigma \rangle = \Tr(\rho^\dagger \sigma)$, then $\mathcal{L}$ can be viewed as an operator acting on this space. 

In this representation, the density matrix $\rho$ is vectorized by the Choi–Jamiołkowski isomorphism \cite{CHOI1975285,JAMIOLKOWSKI1972275}:
\begin{equation}
    \rho = \sum_{ij} \rho_{ij} \ket{i}\!\bra{j} \mapsto \sket{\rho} = \sum_{i, j} \rho_{ij} \ket{i} \otimes \ket{j},
\end{equation}
and super-operators are mapped to operators on the doubled Hilbert space:
\begin{equation}
    A\rho B \mapsto (A\otimes B^T)\sket{\rho}.
\end{equation}

The Liouvillian (Eq.~\eqref{eq:lindbladian_eq}) is represented by:
\begin{align}
    i\L \mapsto &\H :=   H\otimes \bbI - \bbI \otimes H^T   \nonumber \\
    &+ i  \sum_i L_i\otimes L_i^*-\frac{1}{2}\left(L_i^\dagger L_i \otimes \bbI + \bbI \otimes L_i^T L_i^* \right).
    \label{Eq: map_open_bilayer}
\end{align}

\begin{figure}[t]
    \centering
    \includegraphics[scale=0.77]{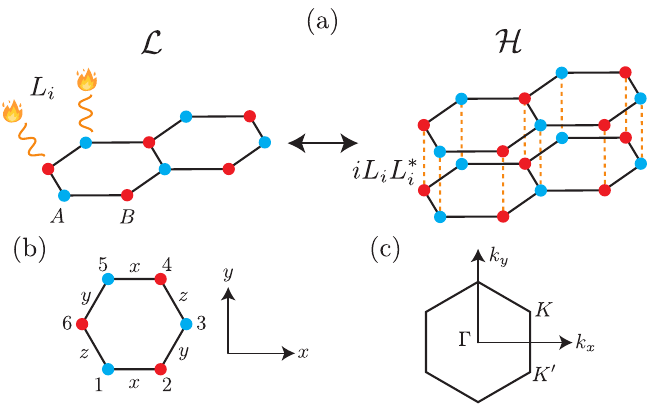}
    \caption{(a) Illustration of the mapping between an open system described by the Liouvillian $\L$ (left) and the non-Hermitian bilayer Hamiltonian $\H$ (right). In the doubled Hilbert space, the jump operators couple the bra and ket spaces. The two sublattices $A$ and $B$ are labeled by blue and red dots, respectively. (b) A hexagonal plaquette in the honeycomb lattice. There are three types of bonds: $x$, $y$, and $z$. (c) The Brillouin zone of the honeycomb lattice.}
    \label{fig: bilayer}
\end{figure}

That is, the Liouvillian $\L$ is mapped to $\mathcal{H}$\footnote{Throughout this paper, $H$ denotes the Hamiltonian part of the open system, while $\H$ denotes the non-Hermitian Hamiltonian in the doubled Hilbert space. More generally, we use calligraphic symbols to denote super-operators.}, a non-Hermitian Hamiltonian living on two copies of the original Hilbert space coupled by the jump operators, as illustrated in Fig.~\ref{fig: bilayer}(a). Indeed, the non-Hermitian terms in $\H$ arise from the jump operators and encode the non-unitary part of the open quantum system's dynamics.

Both frameworks give identical description of the dynamics, since time-evolution of the density matrix is identical, namely:
\begin{equation}
    \dot{\rho} = \L [\rho] \quad \Rightarrow \quad \sket{\dot{\rho}} = -i\mathcal{H} \sket{\rho}.
\end{equation}

In this representation, the Liouvillian spectrum $\{ \lambda_n \}$  is mapped to the energies of $\H$ via $\{E_n = -i\lambda_n\}$, so steady states of the open quantum system $\L \rho_{\text{ss}} = 0$ are mapped to zero-energy eigenstates of $\H$, $\H  \sket{\rho_{\text{ss}}} = 0$. 

Beyond the correspondence between spectra and steady states, the mapping is especially useful for organizing the symmetry structure of Liouvillians. The doubled-Hilbert-space representation provides a natural framework to classify and analyze symmetries of the Liouvillian~\cite{Ryu_classification, L_symmetry, L_symmetry2, L_symmetry3, prosen_PT}. For example, breaking of parity–time ($\P\T$) symmetry in the Liouvillian~\cite{prosen_PT,Huber_PT, PT_open} appears as the emergence of exceptional points (EPs) in the non-Hermitian Hamiltonian, where eigenvalues and eigenvectors coalesce~\cite{El-Ganainy2018,Özdemir2019,RevModPhys.93.015005, PhysRevResearch.4.L042025, Ding_2022}. In Sec.~\ref{Sec: symmetry}, we detail how symmetries of our model are interpreted in the doubled-Hilbert-space formalism.

\section{Exactly solvable dissipative spin orbital model \label{Sec: exactly_solvable}}
We consider a spin–orbital model on the honeycomb lattice with two spin-$1/2$ degrees of freedom per site: a “spin” $\sigma$ and an “orbital” (pseudospin) $\tau$. The Hamiltonian is
\begin{equation}
    H = \frac{J}{4}\sum_{\braket{ij}_\lambda} (\tau_i^\lambda \tau_j^\lambda )\left(\sigma^x_i \sigma_j^x + \sigma^z_i \sigma_j^z \right),
    \label{eq:H_singlelayer}
\end{equation}
where $\braket{ij}_\lambda$ ($\lambda = x, y, z$) denotes the $\lambda$-type bond between two nearest sites $i$ and $j$, as shown in Fig.~\ref{fig: bilayer}(b). The $\sigma$ and $\tau$ variables act on distinct Hilbert spaces on each site and commute with each other.

Our model is a variant of the Yao-Lee (YL) model~\cite{yaolee_original}, which generalizes Kitaev's honeycomb model~\cite{Kitaev_honeycomb} to a spin-orbital setting. The YL Hamiltonian reads:
\begin{equation}
    H_{\text{YL}} = \frac{J}{4}\sum_{\langle ij\rangle_\lambda}
\big(\tau_i^\lambda \tau_j^\lambda\big)
\big(\boldsymbol{\sigma}_i\cdot \boldsymbol{\sigma}_j\big).
\label{eq:YL}
\end{equation}

Compared to YL Hamiltonian (Eq.~\eqref{eq:YL}),
our model (Eq.~\eqref{eq:H_singlelayer}) explicitly breaks the $\text{SU}(2)$ symmetry of the $\sigma$ degree of freedom by dropping the $\sigma_i^y \sigma_j^y$ channel. The remaining continuous symmetry is U$(1)$, as the Hamiltonian is invariant under global rotations $\exp{(i\theta \sum_i \sigma_i^y)}$. This anisotropic choice could be realized by an easy-plane anisotropy in the spin sector and will be essential for the solvability of the Liouvillian.

On top of the Hamiltonian Eq.~\eqref{eq:H_singlelayer}, we apply two jump operators on each site, $\{\sqrt{\gamma} \sigma_i^x, \sqrt{\gamma} \sigma^z_i\}$. The local dephasing terms only act on the $\sigma$-sector and leave the orbital structure unchanged. Following Eq.~\eqref{Eq: map_open_bilayer}, we map the open system to the following non-Hermitian Hamiltonian $\H$ acting on the doubled Hilbert space:
\begin{align}
    \H &=  \frac{J}{4}\sum_{\braket{ij}_\lambda} \tau^\lambda_i \tau^\lambda_j \left(\sigma^x_i \sigma_j^x + \sigma^z_i \sigma_j^z \right) - \tilde{\tau}^\lambda_i \tilde{\tau}^\lambda_j \left(\tilde{\sigma}^x_i \tilde{\sigma}_j^x + \tilde{\sigma}_i^z \tilde{\sigma}_j^z \right) \nonumber \\
    & + i \gamma \sum_i \left(\sigma^x_i \tilde{\sigma}^x_i + \sigma^z_i \tilde{\sigma}^z_i \right) - 2 i \gamma N,
    \label{eq:bilayer_spin}
\end{align}
where $N$ is the number of sites in one layer. Throughout this paper,  the overhead tilde will denote operators that act on the ket copy of the Hilbert space, $\tilde{A} : = \mathbb{I} \otimes A^T$~\cite{CHOI1975285,JAMIOLKOWSKI1972275}.
%up to a purely imaginary constant proportional to $N$, the total number of sites in the bilayer. \yx{This term contributes an overall decay. [further explain]} 
%Here $\{\tau,\sigma\}$ and $\{\tilde{\tau}, \tilde{\sigma}\}$ denote the operators in two layers respectively. 

To solve this model $\H$, we introduce two types of Majorana fermions $c$ and $d$ via the transformation that has been used in the solution of the Yao-Lee model \cite{yaolee_original}:
\begin{align}
    & \sigma_i^\alpha = -\frac{\epsilon^{\alpha \beta \gamma}}{2} i c_i^\beta c_i^\gamma , ~~ \tau_i^\alpha = -\frac{\epsilon^{\alpha \beta \gamma}}{2} i d_i^\beta d_i^\gamma , \nonumber \\
    & \sigma_i^\alpha \tau_i^\beta = i c_i^\alpha d_i^\beta,
    \label{eq:majorana_decomp}
\end{align}
where $\alpha,\beta,\gamma=x,y,z$ and $\epsilon$ is the Levi-Civita symbol. There are six Majoranas per site, forming two triplets $\vec{c}_i$ and $\vec{d}_i$, and similarly for the ket copy.

The Majorana fermions satisfy mutual anticommutation relations, $\{ c_i^\alpha, c_j^\beta \} = \{ d_i^\alpha, d_j^\beta \} = \delta_{ij}^{\alpha \beta}$, $\{c_i^\alpha, d_j^\beta \} = 0$. Similar decompositions are performed on $\tilde{\sigma}$ and $\tilde{\tau}$. Compared to other exactly solvable dissipative models involving Gamma matrices~\cite{arovas_dissipative, Shackleton_dissipative, lucas_sa}, the Majoranas in this model has a more direct connection to the physics variables: namely, the $c$ and $d$ Majoranas are associated with the spin ($\sigma$) and orbital ($\tau$) degrees of freedom, respectively. 

Since above decomposition into Majoranas (Eq.~\eqref{eq:majorana_decomp}) is over-complete~\cite{Kitaev_honeycomb}, we need to project back to the physical space by imposing the local constraint:
\begin{equation}
    D_j = -i c_j^x c_j^y c_j^z d_j^x d_j^y d_j^z,
\end{equation}
where $j$ is the site index. $\widetilde{D_j}$ can be defined similarly for sites in the other layer. A state $\ket{\Psi}$ is only physical if and only if it is an eigenstate of all local projectors with eigenvalue $1$~\cite{yaolee_original}:
\begin{equation}
   \widetilde{D_j} \ket{\Psi} =  D_j \ket{\Psi} = \ket{\Psi}, \, \forall \, j.
\end{equation}

In the Majorana basis, the non-Hermitian Hamiltonian becomes:
\begin{align}
    \H = & \frac{J}{4} \sum_{\braket{ij}_\lambda} \left[ id_i^\lambda d_j^\lambda (ic_i^x c_j^x + i c_i^z c_j^z) - i\tilde{d}_i^\lambda \tilde{d}_j^\lambda(i \tilde{c}_i^x \tilde{c}_j^x + i\tilde{c}_i^z \tilde{c}_j^z) \right] \nonumber \\
    & + \gamma \sum_i (ic_i^y \tilde{c}_i^y ) (i c_i^x \tilde{c}_i^x + ic_i^z \tilde{c}_i^z) -2i\gamma N.
\end{align}

We can further define bond operators:
\begin{equation}
    u_{ij}\equiv id_i^\lambda d_j^\lambda,~~ \tilde{u}_{ij} \equiv i \tilde{d}^\lambda_i \tilde{d}_j^\lambda, ~~ v_i \equiv i c_i^y \tilde{c}_i^y,
    \label{eq:bonds}
\end{equation}
which are defined for all three bond types $\lambda =x,y,z$. The bond operators are static $\mathbb{Z}_2$ gauge fields and have eigenvalues $\pm 1$. Since the bond operators commute with the Hamiltonian $[u_{ij}, \H] = [v_i, \H] = 0$ and among themselves $[u_{ij}, u_{i'j'}] = [u_{ij}, v_i] = 0$, the Hilbert space is split into sectors labeled by their eigenvalues. Note that $\H$ is invariant under a local $\mathbb{Z}_2$ gauge transformation:
\begin{align}
    &c_i^\alpha \rightarrow \Lambda_i c_i^\alpha \, , \, \tilde{c}_i^\alpha \rightarrow \tilde{\Lambda}_i \tilde{c}_i^\alpha  \nonumber \, , \\
    &u_{ij} \rightarrow \Lambda_i u_{ij}\Lambda_j \, , \, \tilde{u}_{ij} \rightarrow \tilde{\Lambda}_i \tilde{u_{ij}} \tilde{\Lambda}_j \, , \, v_i \rightarrow \Lambda_i v_i \tilde{\Lambda}_i,
    \label{eq:local_gauge}
\end{align}
where $\Lambda_i$ and $\tilde{\Lambda}_i$ are $\pm 1$. The gauge-invariant quantities are therefore plaquette fluxes i.e. products of bond operator around plaquettes~\cite{Kitaev_honeycomb, yaolee_original}; see Sec.~\ref{Sec: symmetry} for a more detailed discussion.

Finally, we introduce two species of complex fermions, defined as
\begin{align}
    f_i = \frac{1}{2}(c_i^x + ic_i^z), ~~  \tilde{f}_i = \frac{1}{2}(\tilde{c}_i^x + i \tilde{c}_i^z),
\end{align}
which live on two layers in the doubled Hilbert space. In terms of these fermion operators, the Hamiltonian becomes
\begin{align}
    \H = &  \frac{iJ}{2} \sum_{\braket{ij}_\lambda} u_{ij} \left(f_i^\dagger f_j - f_j^\dagger f_i \right) - \tilde{u}_{ij} \left(\tilde{f}_i^\dagger \tilde{f}_j - \tilde{f}_j^\dagger \tilde{f}_i \right) \nonumber \\
    & + 2\gamma \sum_i v_i(f_i^\dagger \tilde{f}_i - \tilde{f}_i^\dagger f_i)-2i\gamma N.
    \label{eq: bilayer_f}
\end{align}

$\H$ is exactly solvable, since it is \textit{quadratic} in the fermionic operators.
Fixing the bond variables ($u_{ij}$, $\tilde{u}_{ij}$, and $v_{i}$) thus reduces the problem to a quadratic non-Hermitian fermion model on a bilayer honeycomb lattice, which can be diagonalized exactly in each flux sector. The Hamiltonian on the bilayer has many interesting symmetries that can considerably simplify the problem, as we will analyze in the next section.

\section{Symmetry Analysis and Steady States \label{Sec: symmetry}}

In this section, we first analyze the strong and weak symmetries of the dissipative spin-orbital model, Eq~\eqref{eq:bilayer_spin}. We then outline the criterion for choosing flux sectors that contain non-equilibrium steady states (NESS). From this, we show that there are exponentially many NESS in this model, which realizes a dissipative spin liquid.

\subsection{Strong and Weak Symmetry \label{subsec: strong_weak}}
In closed quantum systems, the Hamiltonian $H$ is Hermitian and generates unitary time evolution. Symmetries are represented by unitary operators $U$ that commute with the Hamiltonian, $[U, H]= 0$. This commutation relation implies the existence of a conserved ``charge'' $Q$, namely the generator of this symmetry. $Q$ also commutes with $H$, and its expectation value is therefore conserved in time:
\begin{equation}
    i \frac{d}{dt} \left< Q\right> = \braket{[H, Q]} =  0.
\end{equation}
As a consequence, the Hilbert space decomposes into sectors labeled by eigenvalues of $Q$.  
Under the evolution operator $U(t) = e^{-iHt}$, transitions between different sectors are forbidden, and an initial eigenstate $\ket{\psi_0}$ with eigenvalue $q$ under the charge operator $Q$ will remain an eigenstate of $Q$ with the same eigenvalue:
\begin{equation}
    Q \left( U(t) \ket{\psi_0} \right)= U(t) Q \ket{\psi_0} = q \left(U(t) \ket{\psi_0}\right).
\end{equation}

In open quantum systems, symmetries are defined with respect to the Liouvillian $\L$, and the above notion of symmetry can be generalized. There are two types of symmetries: strong and weak symmetry~\cite{prosen_note,PhysRevA.89.022118}. The interplay between these two types of symmetries, such as spontaneous strong-to-weak symmetry breaking, has drawn lots of recent interest~\cite{SSWSB_wangzhong, SSWSB2,SSWSB3, decoherence_lee,measurement3, SWSSB_1xzd-g9s5}.

When an open quantum system has a \textit{strong} symmetry, there exists an \textit{operator} $S$ that commutes with both the Hamiltonian and the jump operators,
\begin{equation}
    [S, H] = [S, L_i] = 0, \, \forall \, i.
\end{equation}

On the other hand, a \textit{weak} symmetry is satisfied when there exists a \textit{super-operator} $\mathcal{S}$ that commutes with the Liouvillian,
\begin{equation}
    [\mathcal{S}, \mathcal{L}] = 0, \, \mathcal{S}(\rho) = S (\rho) S^\dagger
\end{equation}
for some operator $S$. Note that $[S, L_i] = [S, H]=0$ implies $[\mathcal{S}, \mathcal{L}] = 0$, but not vice versa; a strong symmetry therefore necessarily implies a weak symmetry, but the converse is not true. While both types of symmetries allow for block-diagonalizing the Liouvillian, a strong symmetry has an associated conserved charge, while weak symmetries do not~\cite{prosen_note}.

One can also understand the difference between strong and weak symmetries in the bilayer picture, in which $\mathcal{L}$ is mapped to a non-Hermitian Hamiltonian $\mathcal{H}$ in the doubled space. Both types of symmetries leave $\mathcal{H}$ invariant. However, weak symmetries must act jointly on \textit{both} copies of the doubled Hilbert space to preserve $\mathcal{H}$, whereas strong symmetries only act on \textit{either} copy independently~\cite{Ryu_classification}. Consequently, charges associated with strong symmetries are conserved on \textit{each} layer independently. In contrast, weak symmetry only implies conservation of the \textit{difference} between charges on the two layers, which do not translate to conserved observables of the original single-layer system~\cite{SSWSB_wangzhong}.

The dissipative spin-orbital model we study has an extensive number of strong and weak symmetries. Indeed, similar to Kitaev's honeycomb model~\cite{Kitaev_honeycomb}, it is precisely the model's extensive symmetries that allow for its exact solvability. To understand the strong symmetries, we first consider the dissipation-less limit, $\gamma = 0$. The model reduces to a variant of the Yao-Lee model~\cite{yaolee_original}, whose $\text{SU}(2)$ symmetry is explicitly broken. In this (closed) system, the conserved fluxes are products of bond operators around each hexagon~\cite{yaolee_original}:
\begin{equation}
    W_p = \prod_{\left<jk \right> \in p} u_{jk},
    \label{eq:Wp}
\end{equation}
where $p$ denotes a hexagonal plaquette, and $\left< jk \right>$ runs over each of its six bonds. The operators $W_p$ are gauge-invariant under the local $\mathbb{Z}_2$ transformation in Eq.~\eqref{eq:local_gauge} and commute with the Hamiltonian, $[W_p, H] = 0\, \, \forall \, p$.

In this model, the $W_p$ operators take a particularly simple form in the spin language. Consider a hexagon $p$ whose sites and bonds are labeled as in Fig.~\ref{fig: bilayer}(b). We have:
\begin{align}
    W_p &= u_{12} u_{23}u_{34} u_{45} u_{56} u_{16} \nonumber \\
    &= (i d_1^xd_2^x) (i d_2^y d_3^y) (i d_3^z d_4^z) (i d_4^x d_5^x) (i d_5^y d_6^y)(i d_6^z d_1^z) \nonumber \\
    &= (-i d_1^z d_1^x) (-i d_2^x d_2^y) (-i d_3^y d_3^z) (-i d_4^z d_4^x) (-i d_5^x d_5^y) (-i d_6^y d_6^z) \nonumber \\
    &= \tau_1^y \tau_2^z \tau_3^x \tau_4^y \tau_5^z \tau_6^x,
    \label{eq:Wp_def}
\end{align}
where we have used the anticommutation relation between Majoranas. Thus, although $W_p$ is defined as a product of \textit{bond} operators in the fermionic description, it can be viewed as a product of \textit{on-site} Pauli operators around the hexagon.

\begin{figure}
    \centering
    \includegraphics[width=0.8\linewidth]{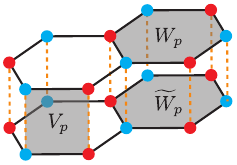}
    \caption{Illustration of the strong and weak $\mathbb{Z}_2$ symmetries of the model in the bilayer framework. The hexagonal plaquettes $W_p$ and $\widetilde{W}_p$ (Eq.~\eqref{eq:Wp}) constitute strong symmetries of the model, while the vertical square plaquettes $V_p$'s (Eq.~\eqref{eq:Vp}) are the weak ($\mathbb{Z}_2$) symmetries.}
    \label{fig:strongweak}
\end{figure}

When dissipation is turned on $(\gamma \neq 0)$, the hexagonal plaquette operators remain conserved. Since $W_p$ act on the orbital sector, while the jump operators $\{\sigma_i^x, \sigma_i^z \}$ act on the spin sector, they commute trivially. Thus
\begin{equation}
    [W_p, L_i] = [W_p, H] = 0,
\end{equation}
and the set of $\{W_p\}$ constitute strong symmetries of the model. In the doubled Hilbert space representation, one can likewise define plaquette operators $\widetilde{W}_p$ acting on the other copy; see Fig.~\ref{fig:strongweak}.

Next, we turn to the model's weak symmetries. Following Refs.~\cite{Shackleton_dissipative, arovas_dissipative, lucas_sa}, we consider the \textit{vertical} plaquettes $V_p$ that link the two layers, as illustrated in Fig.~\ref{fig:strongweak}. Let $i$ and $j$ denote two nearest neighbors on the original layer, connected by a bond of type $\lambda \in \{x, y, z\}$. We define:
\begin{equation}
    V_p = u_{ij} v_i v_j \tilde{u}_{ij}.
    \label{eq:Vp}
\end{equation}

Similar to the hexagonal plaquettes $W_p$, $V_p$ also admits a clean interpretation in the spin variables. The vertical plaquettes can be written as:
\begin{align}
    V_p &= (i d_i^\lambda d_j^\lambda) (i c_i^y \tilde{c}_i^y) (i c_j^y \tilde{c}_j^y) (i \tilde{d}_i^\lambda \tilde{d}_j^\lambda) \nonumber \\
    &=-(i c_i^y d_i^\lambda) (i c_j^y d_j^\lambda) (i \tilde{c_i}^y \tilde{d}_i^\lambda) (i \tilde{c}_j^y \tilde{d}_j^\lambda) \nonumber \\
    &= - \sigma_i^y \sigma_j^y \tau_i^\lambda \tau_j^\lambda \tilde{\sigma}_i^y \tilde{\sigma}_j^y \tilde{\tau}_i^\lambda \tilde{\tau}_j^\lambda.
    \label{eq:Vp_def}
\end{align}

One can similarly check that $[V_p, \H] = 0$, so $V_p$ generates a weak symmetry. In the single-layer picture, this corresponds to a super-operator that performs a unitary conjugation:
\begin{equation}
    \mathcal{S}_{\mathbb{Z}_2} (\cdot) = (\sigma_i^y \sigma_j^y \tau_i^\lambda \tau_j^\lambda) (\cdot) (\sigma_i^y \sigma_j^y \tau_i^\lambda\tau_j^\lambda),
    \label{eq:SZ2_weak}
\end{equation}
which commutes with the Liouvillian super-operator $\mathcal{L}$.

Compared to previously studied exactly solvable DSL models~\cite{arovas_dissipative, Shackleton_dissipative, lucas_sa}, our system hosts an additional, weak $\text{U}(1)$ symmetry under the following super-operator:
\begin{equation}
    \mathcal{S}_{\text{U}(1)} = \exp \left(i \theta \sum_i \left( \sigma_i^y - \tilde{\sigma}_i^y \right) \right),
\end{equation}
where $\theta \in (0, 2\pi)$ is a phase. The weak $U(1)$ symmetry is inherited from the $U(1)$ symmetry of the Yao-Lee variant $H$ (Eq.~\eqref{eq:H_singlelayer}). It is straightforward to verify that $[\mathcal{S}_{\text{U}(1)}, \mathcal{H}] = 0$. Note that since $S = e^{i \theta \sum_i \sigma_i^y}$ does not commute with the jump operators $\{L_i\}$, this is not a strong symmetry.

A direct consequence to this (weak) $U(1)$ symmetry is conservation of the generator $\mathcal{Q} = \sum_i \sigma_i^y - \tilde{\sigma}_i^y$. Interestingly, this can be translated into the conservation of the total number of fermions on the bilayer. Using the Majorana decomposition (Eq.~\eqref{eq:majorana_decomp}), we can write the local fermion number as:
\begin{equation}
   n_i = f_i^\dagger f_i = \frac{1}{4}(c_i^x - i c_i^z)(c_i^x + i c_i^z) = \frac{1}{2} (1 + \sigma_i^y),
\end{equation}
while on the other layer
\begin{equation}
    \tilde{n}_i = \mathbb{I} \otimes (n_i)^T = \mathbb{I} \otimes \left(\frac{1}{2} + \frac{1}{2} \sigma_i^y \right)^T = \frac{1}{2} (1- \tilde{\sigma}_i^y).
\end{equation}

Therefore, conservation of the weak $\text{U}(1)$ charge becomes the weak conservation of the total fermion number, $\sum_i n_i + \tilde{n}_{i}$,. This conservation is precisely why there are no BCS-like pairing terms in the fermionic non-Hermitian Hamiltonian, Eq.~\eqref{eq: bilayer_f}. Instead, the allowed processes are number-conserving: namely, hoppings on each layer, as well as interlayer conversions between the two species of fermions. In Table.~\ref{tab:strongweak}, we summarize the strong and weak symmetries of our model.

\begin{table}[t]
  \centering
  \setlength{\tabcolsep}{6pt}
  \renewcommand{\arraystretch}{1.5}
  \begin{tabular}{ccc}
    \hline\hline
    \textbf{Symmetry} & \textbf{Fermion Picture} & \textbf{Spin Picture} \\
    \hline
    Strong    & $W_p$               & $\tau_1^y \tau_2^z \tau_3^x \tau_4^y \tau_5^z \tau_6^x$ \\
    Strong    & $\widetilde{W}_p$               & $\tilde{\tau}_1^y \tilde{\tau}_2^z \tilde{\tau}_3^x \tilde{\tau}_4^y \tilde{\tau}_5^z \tilde{\tau}_6^x$ \\
    Weak $\mathbb{Z}_2$      & $V_p$               & $-\sigma_i^y \sigma_j^y \tau_i^\lambda \tau_j^\lambda \tilde{\sigma}_i^y \tilde{\sigma}_j^y \tilde{\tau}_i^\lambda \tilde{\tau}_j^\lambda$ \\
    Weak U$(1)$  & $\sum_i (n_i +\tilde{n}_i)$   & $\sum_i \left(\sigma_i^{y} - \tilde{\sigma}_i^y \right)$ \\
    \hline\hline
  \end{tabular}
  \caption{Summary of strong and weak symmetries of the dissipative model. The relevant quantity for each symmetry can written in terms of fermionic and spin variables, both in the bilayer picture. The strong and weak $\mathbb{Z}_2$ symmetries are extensive, whereas the weak U$(1)$ symmetry is global.}
  \label{tab:strongweak}
\end{table}

\subsection{Steady States \label{sec:steadystates}}
Unlike closed (Hermitian) systems, where Lieb's Theorem~\cite{lieb_theorem} dictates that the ground state lies in the sector with zero flux, there are no generic principles for choosing flux configurations in open quantum systems. Nevertheless, strong and weak symmetries still provide useful guidelines for counting non-equilibrium steady states (NESS) of the system.

Let us first analyze the steady states in the two limits, $\gamma = 0$ and $\gamma=\infty$. In the absence of dissipation, the dynamics is unitary under the Hamiltonian in Eq.~\eqref{eq:H_singlelayer}. For a system with $N$ sites and on-site Hilbert space dimension $4$ (spin and orbital), the total Hilbert space dimension is $D=4^N$. If $H$
is nondegenerate, there are $D$ orthogonal energy eigenstates and therefore $D$ linearly independent projectors $\{\ket{\psi_n}\! \bra{\psi_n}\}_{n=1}^{D}$ that all commute with the Hamiltonian. The space of steady states is therefore the $D=4^N$-dimensional space of operators that are diagonal in the energy basis.

In the opposite limit $\gamma  = \infty$, the bilayer Hamiltonian (Eq.~\eqref{eq:bilayer_spin}) reduces to:
\begin{equation}
    \H_{\gamma = \infty} = i \gamma \sum_i \left(\sigma^x_i \tilde{\sigma}^x_i + \sigma^z_i \tilde{\sigma}^z_i \right) - 2i N \gamma.
\end{equation}

Since the sites are decoupled in this limit, it is helpful to instead analyze the single-site Liouvillian $\L_i$:
\begin{equation}
    \L_i(\cdot) = \gamma \left[ \sigma_i^x (\cdot) \sigma_i^x + \sigma_i^z (\cdot) \sigma_i^z - 2(\cdot)  \right].
\end{equation}

One can check that $\L_i(\sigma_i^x) = -2\gamma \sigma_i^x; \L_i (\sigma_i^z) = -2\gamma\sigma_i^z$, while $\L_i(\sigma_i^y) = -4 \gamma\sigma_i^y$, and $\L_i(\mathbb{I}_2) = 0\times \mathbb{I}_2 =0$. Therefore, the only zero mode (steady state) in the $\sigma$ sector is identity, the maximally mixed state, while all traceless components decay. In this limit, steady states factorize as:
\begin{equation}
    \rho_{\text{ss}} = \left(\frac{\mathbb{I}_2}{2}\right)^{\otimes N} \otimes \rho_\tau,
\end{equation}
where $\rho_\tau$ is an \textit{arbitrary} density matrix in the $\tau$ sector. Since the $\tau$-sector Hilbert space dimension is $2^N$, the dimension of the space of steady states is therefore also $(2^N)^2=4^N$.

Beyond the two limits, we need to analyze using symmetries. As shown in Ref.~\cite{prosen_note}, the number of NESS is at least the number of distinct eigenvalues of strong symmetries in the system. Indeed, the Liouvillian is block-diagonalized under left and right actions by strong symmetries, and steady states live in the ``diagonal'' blocks where the left and right labels coincide. In the bilayer representation, this enforces matched hexagonal fluxes, $W_p = \widetilde{W}_p$, for every plaquette $p$, so each steady sector is labeled by a common flux pattern across both layers. Note that each of the hexagonal plaquettes $W_p$ can be $+1$ or $-1$; the only requirement is a matched pattern on the two layers.

Next we analyze the weak symmetries. Following Ref.~\cite{prosen_note}, we will assume that each sector labeled by eigenvalues of the weak symmetries contains at most one NESS. This behavior is typical in local dissipative models~\cite{prosen_note}. Under this assumption, a sector can host an NESS only if it transforms trivially under all weak symmetries (i.e. all weak symmetries have eigenvalue 1). To see this, let $\mathcal{U}$ be a weak symmetry acting as $\mathcal{U}\rho = V \rho V^\dagger$, where $V$ is a unitary operator. If $\rho_{\text{ss}}$ is the unique steady state that satisfies $\L \rho_{\text{ss}} = 0$, then:
\begin{equation}
    0 = \mathcal{U}(\mathcal{L}(\rho_{ss})) = \mathcal{L}(\mathcal{U} \rho_{ss}).
\end{equation}

By uniqueness, $\mathcal{U} \rho_{ss} = u \rho_{ss}$ for some eigenvalue $u$. Taking the trace on both sides yields $\Tr(V \rho_{ss} V^\dagger) = \Tr(\rho_{ss}) = u \Tr(\rho_{ss}) \rightarrow u = 1$. For our model, these constraints select weak-symmetry sectors with a uniform interlayer flux parity $V_p = -1$ on every plaquette. Note that an eigenvalue of $1$ under the weak symmetry in Eq.~\eqref{eq:SZ2_weak} implies $V_p=-1$ due to the minus sign in the definition of $V_p$ in Eq.~\eqref{eq:Vp_def}. It also constraints steady states to sectors with zero relative U$(1)$ charge, $\sum_i \sigma_i^y -\tilde{\sigma}_i^y = 0$. In terms of the fermionic variables, the requirement for zero relative U$(1)$ translates to:
\begin{equation}
    \sum_i n_i +\tilde{n}_i = N,
\end{equation}
i.e. one fermion per site shared across the two layers on average.

We can count the number of NESS following the above analysis. On a honeycomb lattice with $N$ sites and periodic boundary conditions, there are $E = \frac{3}{2}N$ edges. Using Euler's formula for planar graph $V+E-F=2$, where $V$, $E$, $F$ denote the number of vertices, edges, and faces respectively, we see that there are $N/2$ hexagonal plaquettes. The plaquette fluxes satisfy the global constraint $\prod_p W_p = 1$, leaving $\frac{N}{2}-1$ independent fluxes. In addition, there are two independent, non-contractible $\mathbb{Z}_2$ Wilson loops. Hence the number of diagonal sectors under decomposition under the strong symmetries is $2^{N/2-1} \times 2^2 = 2^{N/2+1}$. Imposing the weak-symmetry constraints yields exactly one NESS per (weak symmetry) sector. Therefore, the number of NESS is $2^{N/2+1}$; this may reduce to $2^{N/2-1}$ if global Wilson loops are pinned by boundary conditions. 

The exponentially large number of NESS establishes our model as an example of ``dissipative spin liquid''~\cite{arovas_dissipative,lucas_sa,Shackleton_dissipative}. Dissipative spin liquids are generalizations of quantum spin liquids to the dissipative setting, in which Lindbladian dynamics preserves an emergent gauge structure, enforced by an extensive set of strong symmetries. The macroscopically degenerate manifold of NESS supports nonlocal correlations in the steady states, providing an exactly solvable setting for studying mixed state order~\cite{lucas_sa}. Furthermore, exactly solvable dissipative models admit fractionalized descriptions, where decay modes fall into distinct sectors with parametrically separated relaxation timescales~\cite{Shackleton_dissipative}. Our dissipative Yao-Lee model therefore serves as a solvable, experimentally relevant playground for investigating these exotic non-equilibrium phenomena.

\section{$\P\T$-symmetry breaking and Liouvillian exceptional rings~\label{sec:PT}}
In the previous section, we analyzed strong and weak symmetries to pin down sectors that support non-equilibrium steady states (NESS), namely with $W_p = \widetilde{W}_p, V_p = -1$. However, the transient dynamics outside those sectors remains rich. In fact, the NESS-supporting sectors constitute but an exponentially small fraction of all flux sectors.

In this section, we deliberately break away from sectors that govern long-time dynamics, and study relaxation restricted to a flux sector that does \textit{not} contain NESS. We first analytically compute the single-particle spectrum of the Liouvillian by choosing a translation invariant flux configuration. This amounts to diagonalizing the non-Hermitian Hamiltonian $\H$ in the bilayer picture. We discover that the exceptional points (EPs), at which eigenvalues and eigenstates coalesce, form a ring in momentum space. Eigen-modes inside and outside the ring are $\P\T$ preserving and broken, respectively. We discuss the qualitative consequences of $\P\T$ symmetry breaking on observable relaxation dynamics.

In addition to unitary symmetries, open quantum systems may also exhibit an anti-unitary parity-time ($\P\T$) symmetry. Intuitively, an open quantum system is $\P\T$ symmetric if the combined action of a unitary super-operator $\P$ and anti-unitary super-operator $\T$ leaves the dynamics unchanged, up to a uniform shift in the Liouvillian spectrum.

Interestingly, our dissipative model is $\P\T$ symmetric, according to the criteria by Prosen~\cite{prosen_PT_criteria}. We will take $\P$ to be 
\begin{equation}
    \P(\rho) = U \rho U^\dagger,
\end{equation}
where the unitary $U$ is:
\begin{equation}
    U = \left(\prod_{i=1}^N \sigma_i^y \right) \otimes \mathbb{I}_\tau,
    \label{eq:unitary}
\end{equation}
i.e. $U$ is a parity operator on the $\sigma$ channel and acts trivially on the $\tau$ sector. On the other hand, $\T$ is taken to be conjugation on operators and imaginary numbers:
\begin{equation}
    \T(\rho)\T^{-1} = \rho^* = \rho^T \, \, ; \, \, \T(i)\T^{-1} = -i,
    \label{eq:t_superop}
\end{equation}
where that we have implicitly used the fact that $\rho$ is Hermitian. In Appendix.~\ref{appendix:PT}, we detail the criteria for a Liouvillian to be $\P\T$ symmetric~\cite{prosen_PT, prosen_PT_criteria} and prove that our model satisfies the conditions, for the above choice of $\P$ and $\T$. Therefore, under the mapping $i \L \to \H$, the traceless part of the non-Hermitian Hamiltonian, defined as $\mathcal{H}' = \mathcal{H} - c \mathbb{I}$ with $c = \frac{\text{Tr}(\mathcal{H})}{\text{Tr}(\mathbb{I})}$, is $\P\T$ symmetric in the usual sense for a non-Hermitian Hamiltonian.

Like other symmetries, $\mathcal{PT}$ symmetry can be spontaneously broken. This can be diagnosed directly from the eigenvalue structure of $\H'$. Consider an eigenstate $|\psi\rangle$ of $\H'$, $\H'|\psi\rangle = E|\psi\rangle$.
If $\H'$ is $\mathcal{PT}$ symmetric, $[\mathcal{PT},\H']=0$, then $\mathcal{PT}|\psi\rangle$ is also an eigenstate of $\H'$:
\begin{equation}
    H'(\mathcal{PT}|\psi\rangle) = \mathcal{PT} (H'|\psi\rangle)
    = \mathcal{PT}(E|\psi\rangle)
    = E^*\mathcal{PT}|\psi\rangle),
\end{equation}
where we used the anti-unitarity of $\mathcal{PT}$, which complex-conjugates eigenvalues, in the last step. Thus $\mathcal{PT}$ maps an eigenstate with eigenvalue $E$ to an eigenstate with eigenvalue $E^*$. If the eigenvalues are real, $E=E^*$, it implies $\mathcal{PT}|\psi\rangle \propto |\psi\rangle$, so $|\psi\rangle$ is also an eigenstate of $\mathcal{PT}$ and the mode $\ket{\psi}$ is said to be $\P\T$ preserving. In contrast, if $E$ is complex ($E\neq E^*$), then $|\psi\rangle$ and $\mathcal{PT}|\psi\rangle$ are linearly independent eigenstates with different eigenvalues, so $|\psi\rangle$ cannot be an eigenstate of $\mathcal{PT}$; $\mathcal{PT}$ symmetry is then spontaneously broken for that particular mode~\cite{PhysRevLett.80.5243,10.1063/1.1418246,Bender_2007}.

To best leverage the solvability of our model, we choose a translation-invariant flux configuration, which allows for computing the eigenvalues of $\H'$ in momentum space analytically. Specifically, we fix the intralayer plaquette fluxes to $W_p = \widetilde{W}_p = 1$, and the interlayer plaquettes to $V_p = 1$. Importantly, this sector does \textit{not} host NESS; the NESS reside in sectors where $V_p = -1$. Nonetheless, the choice of this sector is ideal for studying the Liouvillian single-particle spectrum analytically.

In order to realize this flux configuration, we set the bond variables to:
\begin{equation}
    u_{ij}=\tilde{u}_{ij} = v_{i} = 1 \, \, \forall \, i, j.
\end{equation}

We then implement the Fourier transformation by introducing $f_{j\in A} = \frac{1}{\sqrt{N}} \sum_{\vec k} e^{i\vec k\cdot\vec j} a_{\vec k}$ and $f_{j\in B} = \frac{1}{\sqrt{N}} \sum_{\vec k} e^{i\vec k\cdot\vec j} b_{\vec k}$ for sublattices $A$ and $B$ respectively, with analogous expressions for the other fermion species $\tilde{f}$. In momentum space, $\H'$ can be written as
\begin{equation}
    \H' = \sum_{\vec k} \Psi^\dagger(\vec k) h(\vec k) \Psi(\vec k),
\end{equation}
where
\begin{equation}
    h(\vec k) = \frac{1}{2} 
    \begin{pmatrix}
        0 & J i \Delta(\vec{k}) & 4 \gamma & 0 \\
        -J i \Delta(\vec{k})^* & 0 & 0 & 4\gamma \\
        -4\gamma & 0 & 0 &  -J i \Delta(\vec{k})\\
        0 & -4\gamma & J i \Delta(\vec{k})^* & 0
    \end{pmatrix}
    \label{Eq: bloch matrix}
\end{equation}
is the Bloch Hamiltonian, $\Psi(\vec k) = (a_{\vec k}, \, b_{\vec k}, \, \tilde{a}_{\vec k} ,\, \tilde{b}_{\vec k})^T$, and 
\begin{equation}
    \Delta(\vec k) = e^{-ik_x} \left[1 + 2 e^{3ik_x/2} \cos\left(\frac{\sqrt{3}}{2}k_y\right) \right]
\end{equation}
is the structure factor on a honeycomb lattice, where we have set the lattice spacing $a =1$. 

We note that the $\P\T$ symmetry preserves the Bloch Hamiltonian $h(\bm{k})$. Since $\P$ is implemented through a parity operator, and $\T$ is complex conjugation, $\P\T$ acts on the Bloch Hamiltonian as:
\begin{equation}
   (\P\T)h(\mathbf{k})(\P\T)^{-1} =h^*(\mathbf{-k})  = h(\mathbf{k}),
\end{equation}
where we have used the fact that $\Delta(-\mathbf{k}) = \Delta^*(\mathbf{k}).$

Diagonalizing the Bloch Hamiltonian (Eq.~\eqref{Eq: bloch matrix}), we obtain two doubly degenerate bands,
\begin{equation}
    E_{\pm}(\vec k) = \pm \frac{1}{2}\sqrt{J^2|\Delta(\vec k)|^2 - 16 \gamma^2}.
    \label{eq:spectrum_Epm}
\end{equation}

The eigenvalues $E_\pm(\vec{k})$ are purely real when $J^2|\Delta(\vec k)|^2 > 16\gamma^2$, corresponding to $\P\T$ preserving modes and purely imaginary Liouvillian eigenvalues $\lambda(\vec{k})$. When $J^2|\Delta(\vec k)|^2 < 16\gamma^2$, $E_\pm(\vec{k})$ are imaginary and the corresponding Liouvillian modes are decaying in time. The critical points $(k_x^*, k_y^*)$ at which $E=0$ define the boundary between these two qualitatively distinct regimes. At the critical points, the spectrum exhibits a singularity and eigenvalues coalesce in pair, rendering them exceptional points (EPs) of the model~\cite{exceptional_spin_liquid, Ding_2022, nonhermitian_topo_liangfu, RevModPhys.93.015005, PhysRevResearch.4.L042025, Özdemir2019}. The location of EPs can be solved as
\begin{equation}
    1 + 4\cos^2 \left( \frac{\sqrt{3}}{2}k_y^*\right) + 4 \cos\left(\frac{3}{2}k_x^*\right) \cos \left( \frac{\sqrt{3}}{2}k_y^* \right) = 16\frac{\gamma^2}{J^2}.
\end{equation}

In Fig.~\ref{fig: EPs}, we plot the real and imaginary parts of the energy gap, $E_+ - E_-$, over the Brillouin zone for the dissipation strength $\gamma=0.4$, with $J=1$. The set of EPs in momentum space forms an ``exceptional ring'', as shown in the yellow contours. The exceptional rings encircle the $\Gamma$ point $(k_x,k_y)=(0,0)$. Inside the ring, the gap is \emph{purely real}, indicating $\P\T$ preserving modes, while outside the rings it is \emph{purely imaginary}, corresponding to $\P\T$ broken modes.

Note that the $\P\T$ symmetry breaking threshold is different for modes labelled by different momenta. In particular, let $\Delta_{\text{max}} = \max_{\vec{k}} \Delta(\oldbm k) = 3$ denote the maximum of the structure factor over the Brillouin zone.
For $\gamma > \gamma^* = \sqrt{J^2 |\Delta_{\text{max}}(\vec k)|^2 / 16} = 3J/4$, all eigenvalues become entirely imaginary. The momentum-resolved $\P\T$ symmetry breaking is similar to the exceptional rings emerging from interactions and disorders in a Kitaev spin liquid~\cite{exceptional_spin_liquid}. However, the non-Hermitian Hamiltonian in our model arises from an exact mapping from the Liouvillian, as opposed to an effective description incorporating effects of the interactions. Therefore, the dissipation strength $\gamma$ can serve as a tuning parameter for the shape and size of the exceptional rings.

\begin{figure}[t]
    \centering
    \includegraphics[scale=0.315]{gamma_04_EP.pdf}
    \caption{Real (a) and imaginary (b) parts of the energy gap $E_+-E_-$. The exceptional points are highlighted in yellow contours and form a ring in momentum space. Parameters: $\gamma=0.4$, $J = 1$.}
    \label{fig: EPs}
\end{figure}

\begin{figure}[t]
    \centering
    \includegraphics[scale=0.21]{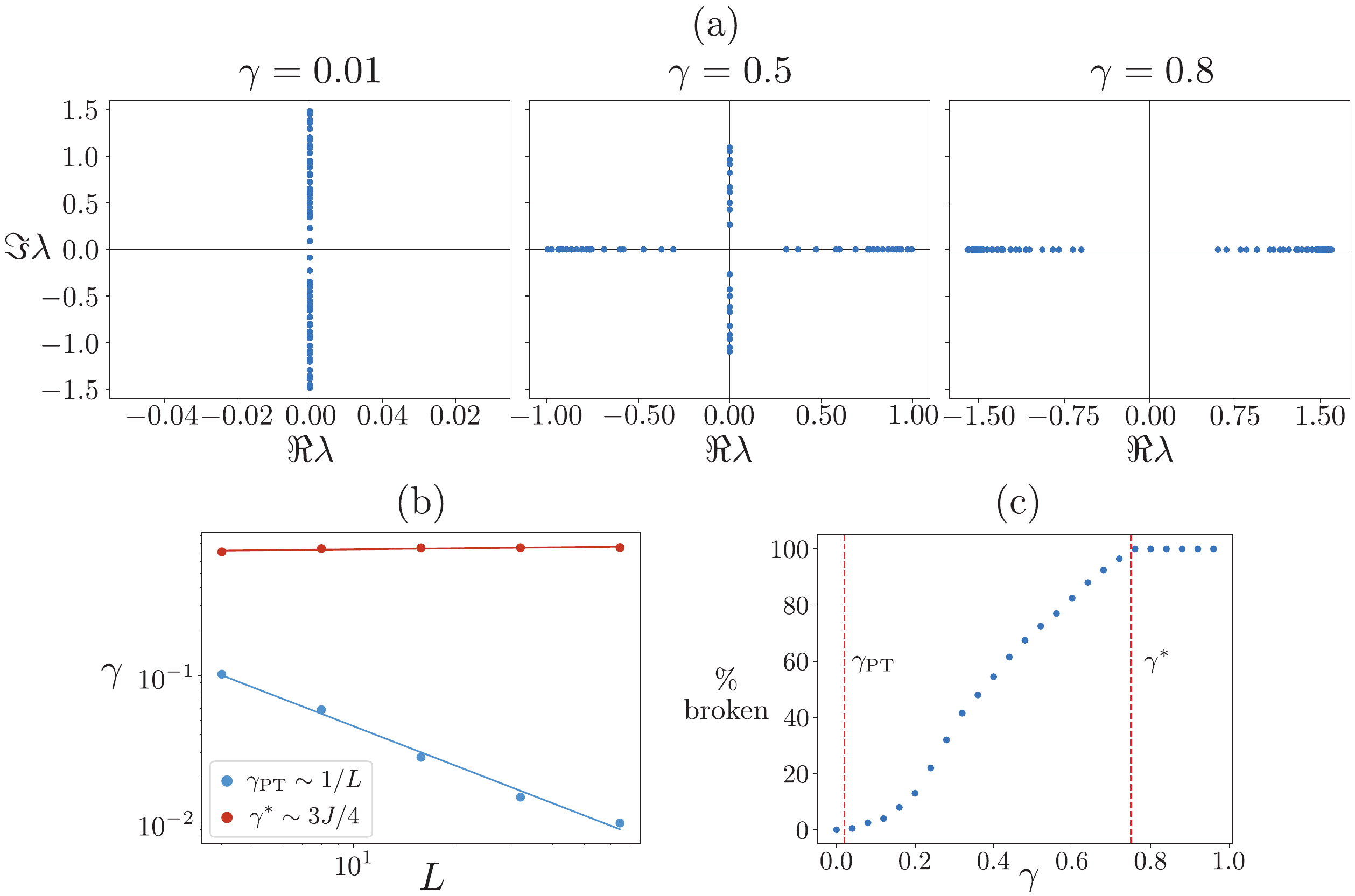}
    \caption{(a) Real-space Liouvillian spectrum for $\gamma=0.01$, $\gamma = 0.5$ and $\gamma = 0.8$. All eigenvalues are shifted by a constant $+\text{Tr}(\L)/\text{Tr}(\mathbb I)$ for plotting purposes. The real-space lattice size is $10\times 10$. (b) Estimated $\gamma_\text{PT}$ and $\gamma^*$ for systems with different linear sizes $L$. The total number of sites is $N=L^2$. $\gamma^*$ approximately remains constant at $3J/4$ for all system sizes, while $\gamma_{\text{PT}}$ decays as $1/L$, as predicted in Ref. \cite{prosen_PT}. Parameters: $L=4,8,16,32,64; J=1$. (c) Fraction of modes that are $\P\T$ broken. The ratio smoothly crosses over from $0$ for $\gamma < \gamma_{\text{PT}}$ to $1$ for $\gamma > \gamma^*$. Parameters: $L=20$, $J=1$.}
    \label{fig: gamma_combined}
\end{figure}

To corroborate our analysis in momentum space, we compute the Liouvillian spectrum $\lambda_n $ in real space, for different dissipation strengths $\gamma=0.01, 0.5,$ and $0.8$. We present the spectrum for a $10\times 10$ lattice in Fig.~\ref{fig: gamma_combined}(a). When there is a small but finite dissipation $\gamma < \gamma_{\text{PT}}$, the eigenvalues of $\H'$ are all real, and therefore $\lambda_n$'s are purely imaginary, corresponding to $\P\T$-preserved modes (left panel of Fig.~\ref{fig: gamma_combined}(a)). As $\gamma$ increases above $\gamma_{\text{PT}}$, some modes become $\P\T$-broken and associate with complex $E_\pm$, while other modes (within the exceptional rings) remain eigenstates of $\P\T$ and correspond to real $E_\pm$. The $\lambda_n$'s are therefore a mixture of real and imaginary values (middle panel of Fig.~\ref{fig: gamma_combined}(a)). Finally, for $\gamma > \gamma^*=3J/4$, all $E_\pm(\vec{k})$ become imaginary, leading to an entirely real Liouvillian spectrum (right panel of Fig.~\ref{fig: gamma_combined}(a)). Each mode therefore decays with a distinct rate, i.e. are all $\P\T$ broken. Our finding of a three-stage crossover as dissipation strength is turned up has also been analyzed in dissipative \textit{fermionic} settings~\cite{dissipative_stark, quantum_quench_drivendissipative}.

The two thresholds, $\gamma_{\text{PT}}$ and $\gamma^*$, correspond to dissipation thresholds above which \textit{some} modes become $\P\T$ broken and \textit{all} modes are $\P\T$ broken, respectively. $\gamma^*$ is only related to the other energy scale in the system, hopping strength $J$, and the lattice geometry. In contrast, $\gamma_{\text{PT}}$ should depend on the system size $N$. We see from Eq.~\eqref{eq:spectrum_Epm} that $\gamma_{\text{PT}}$ is proportional to the \textit{minimum} of $\Delta(\vec{k})$ over the (discrete) Brillouin zone, $\gamma_{\text{PT}} = J/4 \min_{\vec{k}} \Delta(\vec{k})$. In the thermodynamic limit, $\vec{k}$ becomes continuous, and $\min_{\vec{k}} \Delta(\vec{k}) \rightarrow \Delta(K) = 0$, therefore pushing $\gamma_{\text{PT}} \rightarrow 0$. In finite systems, however, there is usually a small but finite $\gamma_{\text{PT}}$. In Fig.~\ref{fig: gamma_combined}(b), we examine the scaling of $\gamma_{\text{PT}}$ and $\gamma^*$ with system size. Our numerical results indicate that $\gamma^*$ saturates at $3J/4$, while $\gamma_{\text{PT}}$ decreases with the system size, consistent with the inverse level-spacing scaling presented in Ref.~\cite{prosen_PT}.

Two remarks are in order regarding the spectrum of $\H'$ and $\L$. First, while we worked with the traceless part $\H'=\H- c \mathbb{I}$, all discussions above apply to the full Hamiltonian \(\H\). Indeed, since $\H$ and $\H'$ are related by a constant offset and share the same set of eigenvectors, the two Hamiltonians have the same $\P\T$ breaking threshold and therefore the same set of EPs. 

Secondly, we emphasize that all results presented above are within a single flux sector $W_p = \widetilde{W}_p = V_p = 1$, chosen for its analytical tractability. The full, many-body spectrum of $\L$ is the union of eigenvalues in all of the (exponentially many) flux sectors. However, since the Liouvillian is block-diagonal under strong and weak symmetries, dynamics is restricted to the initial sector. Using this observation, we could probe the dynamical consequences of $\P\T$ symmetry breaking.

To this end, we first need to prepare an initial density matrix $\rho_0$ that partially lies in the flux sector of interest. Note that no physical density matrix can live \textit{entirely} in the $V_p=1$ sector, due to its traceless-ness. Concretely, we consider two product states, defined as:
\begin{align}
    \ket{\psi_+} := \ket{\uparrow \uparrow \dots \uparrow\uparrow}_y, \\
    \ket{\psi_-} := \ket{\uparrow \downarrow \dots \uparrow\downarrow}_y.
\end{align}
where $\ket{\uparrow}$ and $\ket{\downarrow}$ denote eigenstates of $\sigma_i^y$, respectively. From this, we can define
\begin{equation}
    \ket{\Psi} = \frac{1}{\sqrt{2}} \left( \ket{\psi_+} + \ket{\psi_-}\right),
\end{equation}
and construct the following density matrix:
\begin{align}
    \rho_0 &= \ket{\Psi}\bra{\Psi} \otimes \mathbb{I}_\tau  = \rho_+ + \rho_-,
\end{align}
where
\begin{align}
    \rho_+ = \frac12\left( \ket{\psi_+}\bra{\psi_+} + \ket{\psi_-}\bra{\psi_-}  \right) \otimes \mathbb{I}_\tau, \\
    \rho_- = \frac12\left( \ket{\psi_+}\bra{\psi_-} + \ket{\psi_-}\bra{\psi_+}  \right) \otimes \mathbb{I}_\tau.
\end{align}

From the definition of the weak $\mathbb{Z}_2$ symmetry in terms of the spin variables, Eq.~\eqref{eq:SZ2_weak}, one can easily check that $\rho_+$ has eigenvalue $+1$ under the $\mathbb{Z}_2$ weak symmetry and thus lives entirely in the $W_p = \widetilde{W}_p = 1; V_p = -1$ sector, while $\rho_-$ falls in the $W_p = \widetilde{W}_p = V_p = 1$ sector. $\rho_+$ describes the ``population'', or ``coherence'', between the two states, while $\rho_-$ can be interpreted as the interference. Note that $\text{Tr}(\rho_+)=1$, while $\text{Tr}(\rho_-)=0$, which is consistent with the observation in Sec.~\ref{sec:steadystates} that only $V_p = -1$ hosts steady states, which are of trace 1.

Next we consider a specific observable $M = \sum_i \sigma_i^y$, the magnetization in the $y$-direction. In terms of complex fermions, the magnetization can be written as $M = \sum_i f_i^\dagger f_i - N,$ i.e. it is the fermion occupation number on the original layer, up to a constant. Note that while the \textit{total} fermion number is conserved, the number on each layer can change due to the hopping term in Eq.~\eqref{eq: bilayer_f}. 

Under Liouvillian dynamics, the expectation value of $M$ is:
\begin{align}
    \braket{M(t)} &= \text{Tr}(M \rho(t)) = \text{Tr}(M e^{\L t} \rho_0)  \\
    &= \text{Tr}(M e^{\L t} \rho_+) + \text{Tr}(M e^{\L t} \rho_-).
\end{align}

This expectation value can also be computed in the bilayer picture. Recall that upon vectorization, $M \rightarrow M \otimes \mathbb{I}$, $\rho_0  \rightarrow \sket{\rho_0}$, and $\L \rightarrow -i \H$. Therefore:
\begin{align}
    \braket{M(t)} &= \sbra{M \otimes \mathbb I} e^{-i\H t} \sket{\rho_+} + \sbra{M \otimes \mathbb I} e^{-i\H t} \sket{\rho_-}.
\end{align}

Crucially, since both strong and weak fluxes are conserved quantities, the dynamics remains confined to each sector. This fact allows for computing the two terms separately, as their dynamics are independent. We can extract the qualitative behavior of the expectation value by working in momentum space. Importantly, since both $\rho_+$, $\rho_-$, and $M$ are translation-invariant, only the uniform $(\vec{k}=\Gamma = (0,0))$ mode of the Bloch Hamiltonian contributes to $\braket{M(t)}$. All other momenta average to zero when summed over the Brillouin zone. 

Let us first consider $\sbra{M \otimes \mathbb I} e^{-i\H t} \sket{\rho_+}$. Since the desired flux configuration is $W_p=\widetilde{W}_p=1; V_p=-1$, we choose the bond variables as:
\begin{equation}
    u_{ij}=1; \tilde{u}_{ij} = -1;v_{i} = 1 \, \, \forall \, i, j.
\end{equation}

In momentum space, the Bloch Hamiltonian is now
\begin{equation}
    h(\vec k)^{V_p=-1} = \frac{1}{2} 
    \begin{pmatrix}
        0 & J i \Delta(\vec{k}) & 4 \gamma & 0 \\
        -J i \Delta(\vec{k})^* & 0 & 0 & 4\gamma \\
        -4\gamma & 0 & 0 &  J i \Delta(\vec{k})\\
        0 & -4\gamma & -J i \Delta(\vec{k})^* & 0
    \end{pmatrix}.
\end{equation}
Note the difference between the Bloch Hamiltonian in this sector, compared to one in Eq.~\ref{Eq: bloch matrix}. The spectrum in this sector is thus:
\begin{equation}
    E_\pm^{V_p=-1}(\mathbf{k}) = -2i\gamma + \left(\pm 2i \gamma \pm \frac{J}{2} |\Delta(\vec{k})| \right).
\end{equation}
Crucially, the real and imaginary parts of the spectrum decouple, and modes are either oscillatory, or decaying with a \textit{momentum-independent} rate. Therefore, qualitatively, we expect:
\begin{equation}
   M_{V_p=-1} = \sbra{M \otimes \mathbb I} e^{-i\H t} \sket{\rho_+} \sim C (M_0 +e^{-2\gamma t}).
\end{equation}

Next, we compute $\sbra{M \otimes \mathbb I} e^{-i\H t} \sket{\rho_-}$. From the spectrum Eq.~\eqref{eq:spectrum_Epm} and the fact that $\Delta(\vec{k} = (0,0) ) = 3$, we see that for $\gamma < 3J/4$, the eigenvalues of $h(\vec{k})$ are entirely real, corresponding to oscillatory dynamics. On the other hand, for $\gamma > 3J/4$, $E_{\pm}(\vec{k} = (0,0))$ becomes imaginary, leading to decaying Liouvillian modes. Finally, reinstating the constant shift of Liouvillian spectrum (which we denote as $c_L = \frac{\text{Tr}(\L)}{\text{Tr} (\mathbb{I})}$ and manifests as a rescaling of the observables~\cite{quantum_quench_drivendissipative}), we expect the magnetization to scale as:
\begin{equation}
\braket{M(t)} \sim M_{V_p=-1} + e^{-c_L t} \times
\left\{
\begin{array}{l}
\cos(\sqrt{9J^2-16\gamma^2} t), \gamma < \frac{3}{4} J ; \\
e^{-\sqrt{16\gamma^2-9J^2} t}, \gamma > \frac{3}{4} J,
\end{array}
\right.
\end{equation}

There is a clear, qualitative difference in the dynamics. Specifically, up to a term from the steady state sector and an envelope, $\braket{M(t)}$ remains oscillatory throughout the $\P\T$ mixed phase and transitions to decaying when $\gamma > \gamma^*$. This could in principle be observed in experimental platforms where both the spin-orbital Hamiltonian and local spin dissipations could be engineered. To probe other points in momentum space, one could measure spatially modulated observables.

Our analysis, therefore, provides a qualitative description of the relaxation dynamics for any state prepared within this specific sector. This demonstrates that $\P\T$ symmetry breaking is a genuine dynamical feature of the model, existing alongside the dissipative spin liquid physics (i.e., the NESS) found in other sectors.

\section{Summary and Discussion\label{sec:conclusion}}
In this work, we study a dissipative variant of the Yao-Lee spin-orbital model. By vectorizing density matrices, we map the Liouvillian to a non-Hermitian Hamiltonian acting on two copies of the original Hilbert space. We decompose the spin and orbital operators into Majoranas, followed by rewriting in terms of complex fermions. Since the Liouvillian is quadratic in the fermion operators, the model is exactly solvable. We analyze the model's (extensive number of) strong and weak symmetries and demonstrate that it hosts an exponential large manifold of non-equilibrium steady states.

Furthermore, we break away from sectors that host NESS and analytically compute the Liouvillian eigenvalues in a translationally-invariant symmetry sector. We discover a three-stage behavior of the eigenvalues as the dissipation strength $\gamma$ is increased: first, below a system-size dependent threshold $\gamma_{\text{PT}}$, all eigenvalues are imaginary, corresponding to a $\P\T$ preserving phase. For $\gamma_{\text{PT}}< \gamma < \gamma^*$, the spectrum contains both purely real and purely imaginary eigenvalues, and the system is in a $\P\T$-mixed phase~\cite{quantum_quench_drivendissipative}. Finally, when the dissipation strength is larger than $\gamma^*$, all modes have real eigenvalues and decay with distinct rates, i.e. the system is $\P\T$ broken. We are able to obtain a qualitative (from oscillatory to decaying) crossover of observable dynamics, up to a decaying envelope.

We supplement our analysis with numerical calculations in both real and momentum space. In $k$-space, we discover that the exceptional points form an exceptional ring~\cite{exceptional_spin_liquid}. In real space, we calculate the Liouvillian eigenvalues and demonstrate that the scalings of $\gamma_{\text{PT}}$ and $\gamma^*$ with system size are in agreement with our analysis.

There are a few interesting directions for future exploration. First, in the long-time limit, the relaxation timescale is determined by the Liouvillian gap, which is the slowest non-zero decay rate and governs relaxation dynamics at long times in the system~\cite{liouvillian_gap_def}. More precisely, since every eigenvalue $\lambda_n$ of $\L$ has a non-positive real part~\cite{third_quantization}, we can order the Liouvillian spectrum by its real part as $0= |\Re \lambda_0| < |\Re \lambda_1| < \cdots < |\Re \lambda_n|$. The Liouvillian gap is defined as~\cite{liouvillian_gap_def}
\begin{equation}
    g = |\Re \lambda_1|.
    \label{eq:liouvillian_gap}
\end{equation}

One can analyze ``excitations'' on top of the sectors that host steady states: in particular, the model hosts distinct excitations in the Liouvillian spectrum, corresponding to flipping a vertical plaquette $V_p$, flipping a hexagonal plaquette $W_p$, or creating/removing fermions. Note that since the fermion \textit{number}, instead of parity~\cite{Shackleton_dissipative, arovas_dissipative}, is conserved, there is an extensive number of distinct excitations in the fermionic degree of freedom.

Another quantity that one could study in the long time limit is the purity of steady states, defined as:
\begin{equation}
    P = \text{Tr}(\rho^2).
\end{equation}
From Sec.~\ref{sec:steadystates}, when we analyzed the nature of steady states for $\gamma=0$ and $\gamma=\infty$, we have already seen a hint of this. Without dissipation, the system's steady states are pure states; at very large dissipation, $\rho_{\text{ss}}$ are fully mixed (i.e. identity) in the $\sigma$ sector. Therefore, there might be a \textit{purity transition} of the steady states as the dissipation strength increases~\cite{Huber_PT}.

Finally, it is known from non-Hermitian physics that exceptional points in two dimensions and above have distinct topology compared to those in 1D~\cite{nonhermitian_bandtheory}. While exceptional points in the Liouvillian spectrum have been studied in one-dimensional spin and fermionic systems, their nature has not been elucidated in a two-dimensional, dissipative setting. Our exactly solvable dissipative model, therefore, provides a playground for exploring the topology of Liouvillian exceptional points in a higher dimension.

\begin{acknowledgements}
Z.Q. sincerely acknowledges Chao-Ming Jian for guidance and helpful discussions, particularly regarding the form of the Hamiltonian and jump operators. We thank Junkai Dong and Xuepeng Wang for reading through the manuscript and helpful comments. We acknowledge Peize Ding, Matteo Ippoliti, and Jia-Xin Zhang for discussions. Y.X. thanks the hospitality of Kavli Institute of Theoretical Physics (KITP), where part of this work was completed during the program: Learning the Fine Structure of Quantum Dynamics in Programmable Quantum Matter. This research was supported in part by grant NSF PHY-2309135 to the KITP.
\end{acknowledgements}

\appendix
\section{Conditions for $\P\T$ Symmetric Liouvillian \label{appendix:PT}}
On the super-operator level, a Liouvillian $\mathcal{L}$ is $\P\T$ symmetric if it satisfies~\cite{prosen_PT}:
\begin{equation}
    (\mathcal{{PT}}) \mathcal{L} ' (\mathcal{{PT}})^{-1} = -\mathcal{L}'^\dagger,
\end{equation}
under the combined action of two super-operator $\mathcal{P}$ and $\T$, where
\begin{equation}
    \mathcal{L}' = \mathcal{L} - \frac{\text{Tr}(\mathcal{L})}{\text{Tr}(\mathbb{I})} \mathbb{I}
    \label{eq:traceless_L}
\end{equation}
is the traceless part of the Liouvillian.

The parity operation $\mathcal{P}$ is unitary and squares to identity, $\mathcal{P}^2 = I$. It acts on operators as
\begin{equation}
   \P (\rho) = U \rho U^\dagger,
   \label{eq:parity_superop}
\end{equation}
where $U$ is a unitary operator satisfying $U^2 = 1$. On the other hand, $\mathcal{T}$ is \textit{anti}-unitary, but also squares to identity $\mathcal{T}^2 = I$. A common choice for $\mathcal{T}$ is the complex conjugation (Eq.~\ref{eq:t_superop}), which we adopt in this work.

The above definition is generic on the super-operator level. On the operator level, it turns out that $\L$ is $\P \T$ symmetric if the Hamiltonian $H$ and jump operators $\{L_i \}$ satisfy the following conditions~\cite{prosen_PT_criteria}:
\begin{enumerate}
    \item $\{L_m, L_m^\dagger\} = c_m \mathbb{I}$ for some real $c_m$.
    \item $U H^* U^\dagger = H$, where $H$ is the Hamiltonian, and $U$ is the unitary operator that implements parity as in Eq.~\eqref{eq:parity_superop}.
    \item The set of jump operations $\{ L_i \}$ is \textit{closed} under $\P\T$ symmetry. In particular:
\begin{equation}
    U L_i U^\dagger =- \sum_j V_{ij} L_j^\dagger,
\end{equation}
where $V$ is a unitary matrix. Intuitively, the negative sign comes from the anticommutator in the Lindbladian, and this condition ensures that the jump terms $\sum_i L_i L_i^\dagger$ in $\L$ remain invariant under $\P\T$ symmetry.
\end{enumerate}

Our dissipative spin-orbital model satisfies all of the above conditions. The first condition is immediately met, as $\{\sigma_i^x, (\sigma_i^x)^\dagger \} = \{\sigma_i^z, (\sigma_i^z)^\dagger \} = 2 \mathbb{I}$ for all sites $i$. 
Under conjugation by the unitary $U$ defined in Eq.~\eqref{eq:unitary}, the Hamiltonian (Eq.~\eqref{eq:H_singlelayer}) (which is real, $H = H^*$) remains invariant. While conjugation by $U$ flips the sign of individual $\sigma_i^{x}$ and $\sigma_i^z$, $H$ only contains two-body terms $\sigma_i^x \sigma_j^x$ and $\sigma_i^z \sigma_j^z$ in the spin channel, and therefore condition (2) is satisfied. Finally, since $U \sigma_i^z U^\dagger = -\sigma_i^z$, $U \sigma_i^x U^\dagger = -\sigma_i^x$, and the jump operators are Hermitian, $L_i = L_i^\dagger$, condition (3) is satisfied, with $V = \mathbb{I}$. Therefore the Liouvillian in our model is indeed $\P\T$ symmetric.

\bibliography{refs}

\end{document}